\documentclass[twocolumn,superscriptaddress]{revtex4}
\usepackage{amssymb}
\usepackage{amsmath}
\usepackage{graphicx}
\usepackage{subfigure}
\usepackage{natbib}
\usepackage{epsfig}
\usepackage{amsfonts}
\usepackage{mathrsfs}
\usepackage{ulem}
\usepackage{color}
\usepackage[toc,page,title,titletoc,header]{appendix}
\usepackage{multirow}

\begin{document}

\title{Topological incommensurate magnetization plateaus in quasi-periodic quantum spin chains}
\author{Hai-Ping Hu}
\affiliation{
Beijing National Laboratory for Condensed Matter Physics, Institute of Physics, Chinese Academy of Sciences, Beijing 100190, China}

\author{Chen Cheng}
\affiliation{
Center for Interdisciplinary Studies $\&$ Key Laboratory for Magnetism and Magnetic Materials of the MoE, Lanzhou University, Lanzhou 730000, China}

\author{Hong-Gang Luo}
\affiliation{
Center for Interdisciplinary Studies $\&$ Key Laboratory for Magnetism and Magnetic Materials of the MoE, Lanzhou University, Lanzhou 730000, China}
\affiliation{
Beijing Computational Science Research Center, Beijing 100084, China}

\author{Shu Chen}
\thanks{Corresponding author, schen@iphy.ac.cn}
\affiliation{
Beijing National Laboratory for Condensed Matter Physics, Institute of Physics, Chinese Academy of Sciences, Beijing 100190, China}
\affiliation{
Collaborative Innovation Center of Quantum Matter, Beijing, China}
%\affiliation{ Correspondence to schen@aphy.iphy.ac.cn}

\maketitle
%\begin{abstract}
{\bf
Uncovering topologically nontrivial states in nature is an intriguing and important issue in recent years. While most studies are based on the topological band insulators, the topological state in strongly correlated low-dimensional systems has not been extensively explored due to the failure of direct explanation from the topological band insulator theory on such systems and the origin of the topological property is unclear. Here we report the theoretical discovery of strongly correlated topological states in quasi-periodic Heisenberg spin chain systems corresponding to a series of incommensurate magnetization plateaus under the presence of the magnetic field, which are uniquely determined by the quasi-periodic structure of exchange couplings. The topological features of plateau states are demonstrated by the existence of non-trivial spin-flip edge excitations, which can be well characterized by nonzero topological invariants defined in a two-dimensional parameter space. Furthermore, we demonstrate that the topological invariant of the plateau state can be read out from a generalized Streda formula and the spin-flip excitation spectrum exhibits a similar structure of the Hofstadter's butterfly spectrum for the two-dimensional quantum Hall system on a lattice.
~\\
}

Since the discovery of topological insulators \cite{KM1,KM2,BHZ} nearly ten years ago, topological states have attracted great interests in condensed matter physics both theoretically and experimentally \cite{TIRMP1,TIRMP2,symmetrycla}. A hallmark feature of these exotic phases is the appearance of gapless edge states which is robust against local perturbations as long as the bulk gap is not closed.
%protected by band gaps in the spectrum.
To characterize these states, global topological invariants rather than local order parameters should be introduced. Although topological states based on band theory have been well understood, till today the goal of searching topological states in strongly correlated systems remains fascinating and challenging \cite{wangz1,wangz2,etang,lfid,oshi,xchen1,xchen2,pollmann}.

While most of previous studies on topological states focus on either two-dimensional (2D) or three-dimensional materials, recent researches on one-dimensional (1D) periodic and quasi-periodic systems have revealed these systems exhibit non-trivial topological properties \cite{langlijun,kraus1} due to a nontrivial link between these 1D systems and 2D topological insulators \cite{langlijun,kraus1,Mei,kraus3,lilinghu,aamodel1,aamodel2}.
Experimentally, using the propagation of light in photonic waveguides, topologically protected boundary states \cite{kraus1} and phase transition are also observed \cite{kraus3}. The 1D quasi-periodic crystal can be viewed as the simplest realization of a topologically nontrivial insulator. A crucial question is: for more general 1D systems which inevitably suffer from strong quantum fluctuations, can topological states induced by the quasi-periodic modulation survive in the strong correlated regime? If these states exist, they are undoubtedly the strongly correlated topological states being persistently sought by condensed matter physicists. The existence of powerful numerical methods for 1D correlated systems, e.g., the density matrix renormalization group (DMRG) method, permits us to explore novel correlated topological states in a numerically exact way.

In this paper, we investigate the paradigmatic strongly correlated model, i.e., quantum Heisenberg model on a 1D quasi-periodic lattice. We report the findings of a series of non-trivial incommensurate magnetization plateaus as consequence of the existence of large excitation gaps in quasi-periodic quantum spin chains. Quite surprisingly, these incommensurate plateaus will approach to specific non-trivial irrational values which are uniquely determined by the quasi-periodic modulation parameter in thermodynamic limit. The nontrivial topological properties of the incommensurate plateaus are unveiled by using two independent methods, i.e., calculating the edge excitations and topological invariants, both of which are well established and have been widely adopted in the study of topological states \cite{TIRMP1,TIRMP2,shengdn}. Under open boundary conditions (OBC), we find that these non-trivial plateaus can host robust edge spin-flip excitations which connect the lower and upper excitation bands. Different plateau states can be well characterized by topologically invariant Chern numbers, which are defined in a 2D parameter space and are related to the height of plateaus via a generalized Streda formula. It is interesting that the spin-flip excitation spectrum of the quasi-periodic Heisenberg model exhibits a butterfly-like structure, which resembles the Hofstadter spectrum of the 2D quantum hall system.
%Thus our work unifies the quantized plateaus, corresponding non-trivial states for one dimensional quasi-periodic spin systems with quantum Hall conductivity %plateaus and gapless edge states in quantum hall systems in the strongly correlated topological insulator scheme.

\section*{Results}
%\subsection{Incommensurate plateaus.}
{\bf Incommensurate plateaus.}
We consider a general Heisenberg spin-$S$ chain with quasi-periodic geometry which is described by
\begin{equation}
H=\sum_i J_i(S_i^x  S_{i+1}^x+S_i^y  S_{i+1}^y+ S_i^z  S_{i+1}^z) \label{xxx}
\end{equation}
with
\begin{equation}
J_i = J [ 1- \lambda \cos(2\pi\alpha i+\delta)]  \label{Ji}
\end{equation}
where we take the quasi-periodic modulation parameter $\alpha\in(0,1)$  as an irrational number. The exchange strength $J_i$ is quasi-periodic with modulation strength $\lambda$ and phase factor $\delta$. The special case with $\lambda=0$ reduces the Hamiltonian to the homogenous Heisenberg model. In this work, we focus on the anti-ferromagnetic (AFM) couplings, i.e., $J>0$ and $|\lambda|<1$. For convenience, $J=1$ is taken as the unit of energy.
%Note the exchange coupling $J_i$ is invariant under the following transformation $(\alpha,\delta)\rightarrow(\alpha*\equiv 1-\alpha,-\delta)$.
First we study the magnetization process under magnetic field $h$ which couples to the $z$ component of spins \cite{mp1,mp2}. The magnetization per spin is defined as $m_z =S_z/L$ with $S_z=\sum_i^L S_i^z$ being the $z$ component of the total spin and $L$ denoting the lattice size. In Fig.1 we demonstrate the magnetization curves of quasi-periodic spin-$1/2$ Heisenberg chain with three typical irrational numbers (a) $\alpha=\frac{\sqrt{2}-1}{2}$, (b) $\alpha=\frac{\sqrt{3}-1}{2}$, and (c) $\alpha=\frac{\sqrt{5}-1}{2}$ for different chain length $L=50$, $L=100$, and $L=200$ under periodic boundary condition (PBC) calculated by the DMRG method. As the modulation of exchange coupling is quasi-periodic, there is a mismatch between the N-th and the 1st bond as we apply this periodic boundary condition, which contributes to the finite size effect. Except for the trivial plateaus at $m_p=\pm\frac{1}{2}$ corresponding to totally polarization and ignorable minor plateaus due to finite system size, we can clearly observe the emergence of a series of unexpected large plateaus. For different chain lengths with a specific quasi-period $\alpha$, the widths and positions of the plateaus are nearly unchanged. Take $\alpha=\frac{\sqrt{3}-1}{2}$ as an example. From bottom to top, all the plateaus are $-\frac{1}{2}+(0, \frac{18}{50}, \frac{19}{50}, \frac{31}{50}, \frac{32}{50}, 1)$ for $L=50$,
$-\frac{1}{2}+(0, \frac{36}{100}, \frac{37}{100}, \frac{63}{100}, \frac{64}{100}, 1)$ for $L=100$, and
$-\frac{1}{2}+(0, \frac{73}{200}, \frac{74}{200}, \frac{126}{200}, \frac{127}{200}, 1)$ for $L=200$.
As shown in Fig.1, the positions of these magnetization plateaus are fairly close to some $\alpha$-dependent values, e.g., $m_z\approx-S+(\alpha,2\alpha,1-2\alpha,1-\alpha)$ for $\alpha=\frac{\sqrt{2}-1}{2}$, and $m_z\approx-S+(\alpha,1-\alpha)$ for $\alpha=\frac{\sqrt{3}-1}{2}$ and $\alpha=\frac{\sqrt{5}-1}{2}$, where $S=1/2$.
\begin{figure}
\includegraphics[width=3.8in]{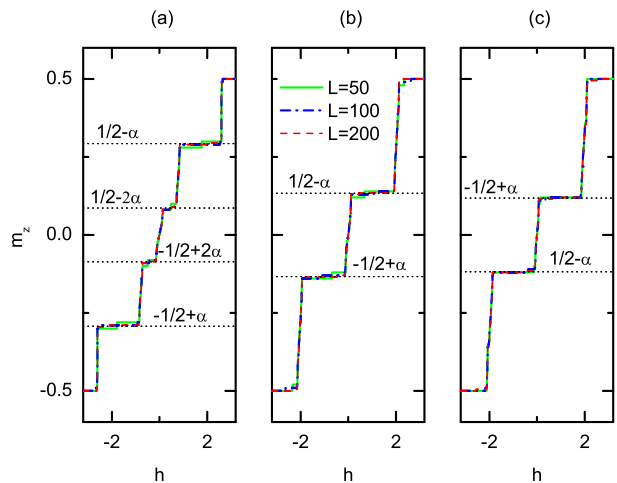}
\caption{(Color online) Magnetization versus $h$ for systems with $\lambda=0.8$, $\delta=0$ and different chain lengths under PBC. From left to right, (a) $\alpha=\frac{\sqrt{2}-1}{2}$, (b) $\alpha=\frac{\sqrt{3}-1}{2}$, (c) $\alpha=\frac{\sqrt{5}-1}{2}$. $L=50$, $100$ and $200$ are represented by solid (green), dashed dotted (blue) and dashed (red) lines, respectively. The horizontal dashed (black) lines denote the specific $\alpha$-dependent values discussed in the main text.}
\end{figure}

The above incommensurate magnetization plateaus for finite-size systems, which approximate to the special values $-S+(\alpha,2\alpha,1-2\alpha,1-\alpha)$, will tend exactly to these values in thermodynamic limit $L\rightarrow\infty$. For brevity, we mark these irrational plateaus from bottom to top as $P_{\alpha}$,$P_{2\alpha}$,$P_{-2\alpha}$,$P_{-\alpha}$. Take a specific plateau $P_{\alpha}$ as an example. As $\alpha$ is irrational, $L\alpha$ is not an integer. Denote $N_{l}$ and $N_{u}$ as the nearest lower and upper bound integer with $N_{l}<L\alpha<N_{u}$. In our spin model, magnetization for the system with $N_{l}$ or $N_{u}$ up spins is $m_{l}\equiv -1/2 + N_{l}/L$ or $m_{u}\equiv-1/2 + N_{u}/L$, respectively. Obviously, we have $m_{l}<P_{\alpha}<m_{u}$. For a finite chain with the length $L$, our DMRG results show that the magnetization plateau is located at either $m_{l}$ or $m_{u}$ as illustrated in Fig.2. When the length $L$ increases, positions of magnetization plateaus exhibit damped oscillations. For different $\alpha$, e.g., (a) $\alpha=\frac{\sqrt{2}-1}{2}$ and (b) $\alpha=\frac{\sqrt{3}-1}{2}$, the positions of plateaus will definitely tend to $P_{\alpha}=-1/2+\alpha$ as $lim_{L\rightarrow\infty}m_{l}=lim_{L\rightarrow\infty}m_{u}=P_{\alpha}$. In thermodynamic limit, these incommensurate plateaus will eventually evolve into irrational magnetization plateaus $P_{\alpha}$. Further we define $\alpha_{l}\equiv\frac{N_{l}}{L}$ and $\alpha_{u}\equiv\frac{N_{u}}{L}$, where $\alpha_{l}<\alpha<\alpha_{u}$ and $\lim_{L\rightarrow\infty}\alpha_{u}=\lim_{L\rightarrow\infty}\alpha_{l}=\alpha$. If we consider these rational $\alpha_{l}$ and $\alpha_{u}$ modulation of spin chains with length $L$, commensurate plateaus at $m_{l}$ and $m_{u}$ appear \cite{OYA}. Based on the above discussion, we can conclude that magnetization plateaus for the quasi-periodic spin chain are totally determined by the irrational modulation parameter $\alpha$.
\begin{figure}
\includegraphics[width=3.8in]{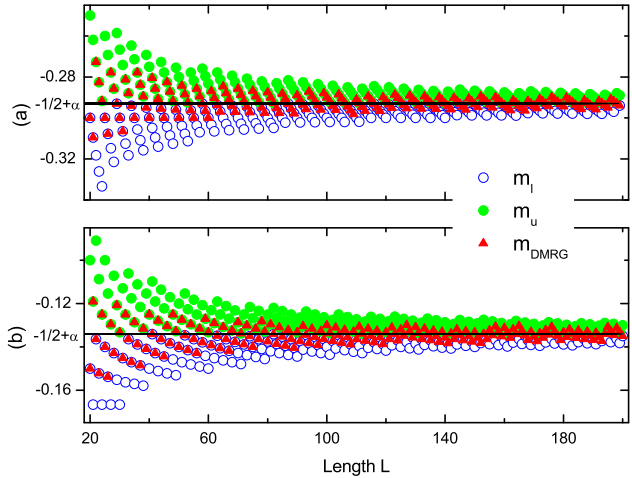}
\caption{(Color online) Magnetization plateaus of incommensurate AFM spin-1/2 Heisenberg chains versus the chain length $L$ for systems with $\delta=0$, $\lambda=0.8$ and different $\alpha$ under PBC. (a) $P_{\alpha}$ plateau for $\alpha=\frac{\sqrt{2}-1}{2}$, $h=-2.2$; (b) $P_{\alpha}$ plateau for $\alpha=\frac{\sqrt{3}-1}{2}$, $h=-1$. Here $m_{DMRG}$ represents magnetization plateau calculated by the DMRG method. The black guidelines denote the corresponding irrational values of magnetization plateaus in thermodynamic limit.}
\end{figure}

%\subsection{Topological properties for plateau states.}
{\bf Topological edge excitations for plateau states.}
The emergence of magnetization plateaus reveals that there exist finite excitation gaps. The size of the gap is proportional to the width of the plateau. When the phase $\delta$ is adiabatically changed, we find that these non-trivial plateaus can host continuous edge spin-flip excitations which connect the lower and upper excitation bands just like those for general topological states under OBC. The adiabatical evolution of phase $\delta$ can be regarded as a generalized Thouless charge pump \cite{thouless1} with $H(\delta+2\pi)=H(\delta)$. Define the spin-flip excitation energy as $\Delta E_{S_z}=E_{S_z+1}-E_{S_z}$. The spin distribution for this excitation is $\Delta\rho_{S_z}=\rho_{S_z+1}-\rho_{S_z}$, where $\rho_{S_z}(i)=\langle\psi| S_i^z|\psi\rangle$ with $\psi$ and $E_{S_z}$ the ground-state wave function and energy in the total $S_z$ subspace. In the following, we focus on the case of $\alpha=\frac{\sqrt{2}-1}{2}$, for which there exist five excitation bands separated by four large excitation gaps. From bottom to top, these gaps lead to magnetization plateaus $P_{\alpha}$, $P_{2\alpha}$, $P_{-2\alpha}$ and $P_{-\alpha}$ in magnetization curves. The adjacent bands are connected by several in-gap excitation modes with adiabatical change of $\delta$ as shown in Fig.3(a). Further, these in-gap modes under OBC are edge modes considering the spin-flip distributions are mainly localized at two ends of the chain as illustrated in Fig.3(b) and Fig.3(c). Our numerical results show that once the edge modes touch the bulk band, the distributions will change side. With no gap closing on path, the cyclical change of $\delta$ leads to windings of edge modes around excitation gaps \cite{hatsugai}. The winding numbers for these four non-trivial plateau states are $1$, $2$, $-2$, and $-1$, respectively.
\begin{figure}
\includegraphics[width=3.8in]{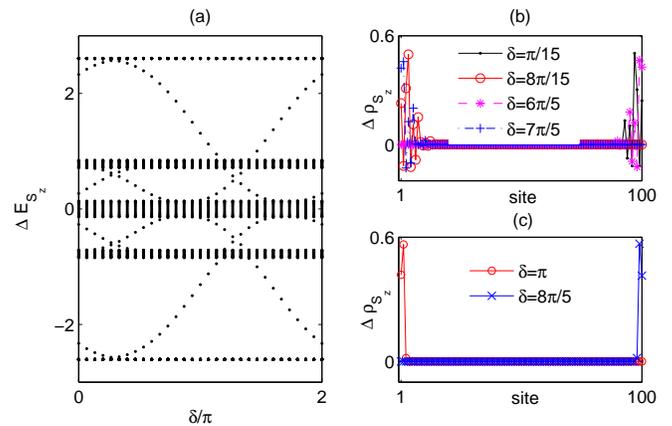}
\caption{(Color online) (a) The spin-flip excitation spectrum $\Delta E_{S_z}$ with respect to the phase factor $\delta$ for spin-$\frac{1}{2}$ AFM  Heisenberg chain with $\alpha=\frac{\sqrt{2}-1}{2}$, $\lambda=0.8$ and $L=100$ under OBC. The four large excitation gaps support the magnetization plateaus $P_{\alpha}$, $P_{2\alpha}$, $P_{-2\alpha}$ and $P_{-\alpha}$ from bottom to top. (b) and (c) Distributions $\Delta\rho_{S_z}$ of in-gap spin-flip excitation modes for plateaus $P_{2\alpha}$ and $P_{\alpha}$, respectively. For each in-gap mode, $S_z=m_{l}L$.}
\end{figure}

%\subsection{Quasi-periodic spin-$1$ chains}
{\bf Quasi-periodic spin-$1$ chains.}
We have demonstrated the non-trivial edge excitations of incommensurate plateau states for quasi-periodic spin-$\frac{1}{2}$ chains. In this part, we extend the study to the quasi-periodic spin-$1$ Heisenberg model. As has been noticed by Haldane \cite{haldane1,haldane2} based on the low-energy effective field theory, half-odd-integer and integer spin chain systems exhibit quite different behaviors. The well known Haldane's conjecture for homogeneous Heisenberg AFM model states that the low energy excitation is gapless for half-odd-integer spins while gapped for integer spins. Though the extension from the spin-$\frac{1}{2}$ to spin-$1$ model is not straightforward, topological states induced by the quasi-periodic geometry exhibit some general behaviors. The magnetization curves are uniquely determined by the quasi-periodic modulation parameter $\alpha$. Take $\alpha=\frac{\sqrt{3}-1}{2}$ for example. The magnetization process is different from the spin-$\frac{1}{2}$ model with the appearance of more incommensurate magnetization plateaus by our DMRG calculations. As shown in Fig.4(a), for the finite spin-1 chain with $L=100$ under PBC, the magnetization plateau emerges at $-1+(0; \frac{36}{100}, \frac{37}{100}; \frac{73}{100}, \frac{74}{100}; 1; 2-\frac{74}{100}, 2-\frac{73}{100}; 2-\frac{37}{100}, 2-\frac{36}{100}; 2)$ from bottom to top. The existence of the middle zero-plateau is a reminiscence of Haldane gap \cite{haldane1}. Other nontrivial incommensurate plateaus will approach to $-1+(\alpha; 2\alpha; 2-2\alpha; 2-\alpha)$ in thermodynamical limit. Denote these plateaus as $P_{\alpha}$, $P_{2\alpha}$, $P_{-2\alpha}$, $P_{-\alpha}$. In Fig.4(b) we show the spin-flip excitations $\Delta E_{S_z}$ with respect to the phase factor $\delta$. For these non-trivial plateau states, there exist continuous spin-flip excitations which connect the lower and upper excitation bands.
\begin{figure}
\includegraphics[width=3.8in]{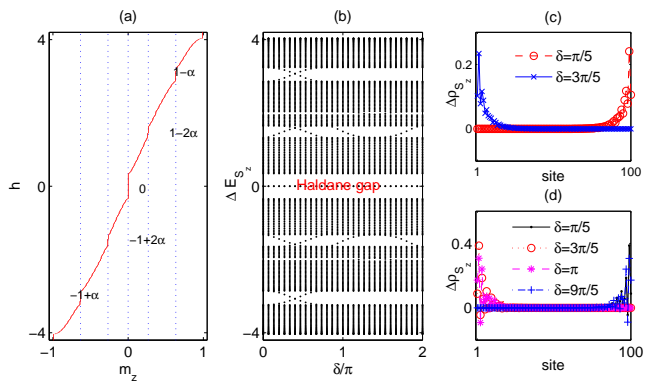}
\caption{(Color online) (a) Magnetization curves for AFM spin-1 Heisenberg model with  $\alpha=\frac{\sqrt{3}-1}{2}$, $\lambda=0.3$ and $L=100$ under PBC. The blue dotted lines are guidelines for the incommensurate plateaus. (b) Spin-flip excitation spectrum $\Delta E_{S_z}$ with respect to phase $\delta$ for spin-$1$ AFM  Heisenberg chains with $\alpha=\frac{\sqrt{3}-1}{2}$ under OBC. The five large excitation gaps correspond to magnetization plateaus $P_{\alpha}$, $P_{2\alpha}$, ``Haldane" 0-plateau, $P_{-2\alpha}$ and $P_{-\alpha}$ from bottom to top. (c) and (d) Distributions $\Delta\rho_{S_z}$ of in-gap spin-flip excitation modes for plateaus $P_{-\alpha}$ and $P_{-2\alpha}$, respectively. For each in-gap mode, $S_z=m_{l}L$.}
\end{figure}

The corresponding spin-flip distributions of in-gap modes for plateaus $P_{-\alpha}$ and $P_{-2\alpha}$ as illustrated in Fig.4(c) and Fig.4(d) clearly demonstrate that these in-gap modes are edge modes. Touching with bulk bands changes the side of edge modes. From bottom to top, the winding numbers for these four incommensurate plateaus are $1$, $2$, $-2$ and $1$, respectively.
On the contrary, the edge modes in the Haldane gap do not connect different bands and the winding number is zero.

%\subsection{Topological invariants}
{\bf Topological invariants.}
According to the bulk-edge correspondence for topological states, the existence of nontrivial edge states is generally attributed to the non-trivial topology of bulk states \cite{hatsugai}. Such a correspondence holds true even for topologically nontrivial interacting systems \cite{wen1,wen2,zhu2,xuz}.
As both spin-$\frac{1}{2}$ and spin-1 quasi-periodic chains display nontrivial edge excitations, we can summarize that, for a general spin-$S$ chain with quasi-periodic modulation parameter $\alpha$, the nontrivial magnetization plateaus in thermodynamic limit should appear at the following specific values: $\pm(S-\alpha,S-2\alpha,...,S-n\alpha...)$ as long as $S-n\alpha>0$. The differences for various non-trivial plateau states are the number of edge states and the winding pattern with respect to the phase factor $\delta$. A natural question is how to define topological invariants to characterize different plateau states. As $\alpha$ totally determines the position of plateaus, the adiabatical evolution of $\delta$ produces a family of systems with quite similar magnetization curves, we can define the topological invariants for the plateau states associated with excitation gaps in a 2D manifold spanned by $(\theta,\delta)$ \cite{thouless1,thouless2,niuqian} where $\theta$ is the twist angle introduced by applying the twist boundary condition to the many body wave function $\psi$. For an arbitrary site $j$, the twist boundary condition is $\psi(j+L,\delta)=e^{i\theta}\psi(j,\delta)$, which has been widely used in spin systems \cite{hatsugaiberryphase,Shastry}. The Chern number is defined as the integral of Berry curvature \cite{thouless2,niuqian} $F(\theta,\delta)$ on the 2D manifold given by
\begin{eqnarray}
&& C = \frac{1}{2\pi}\int d\theta d\delta F(\theta,\delta), \nonumber \\
&& F(\theta,\delta)= Im(\langle\frac{\partial\psi}{\partial\delta}|\frac{\partial\psi}{\partial\theta}\rangle-
\langle\frac{\partial\psi}{\partial\theta}|\frac{\partial\psi}{\partial\delta}\rangle) .\label{Chernnumber}
\end{eqnarray}
In table.1 we list the Chern numbers for several different magnetization plateaus that we have calculated.
\begin{table}
\centering
\caption{Chern numbers for plateau states of quasi-periodic Heisenberg spin chains (here $\alpha*=1-\alpha$)}\label{cherntable}
\begin{tabular}{|c|c|c|c|c|c|c|}
\hline
\multirow{2}{*}{quasi-period-$\alpha$} & \multicolumn{2}{|c|}{$\alpha=\frac{\sqrt{2}-1}{2}$} & \multicolumn{2}{|c|}{$\alpha=\frac{\sqrt{3}-1}{2}$} & \multicolumn{2}{|c|}{$\alpha=\frac{\sqrt{5}-1}{2}$}\\
\cline{2-7}
& $m_p$ & $C_p$ & $m_p$ & $C_p$ & $m_p$ & $C_p$\\
\hline
\multirow{2}{*}{spin-$\frac{1}{2}$} & $\pm(S-\alpha)$ & $\mp1$ & \multirow{2}{*}{$\pm(S-\alpha)$} & \multirow{2}{*}{$\mp1$} & \multirow{2}{*}{$\pm(S-\alpha*)$} &\multirow{2}{*}{$\pm1$} \\
\cline{2-3}
 & $\pm(S-2\alpha)$ & $\mp2$ & & & &\\
\hline
\multirow{4}{*}{spin-1} & $\pm(S-\alpha)$ & $\mp1$ & \multirow{2}{*}{$\pm(S-\alpha)$} & \multirow{2}{*}{$\mp1$} & \multirow{2}{*}{$\pm(S-\alpha*)$} &\multirow{2}{*}{$\pm1$} \\
\cline{2-3}
 & $\pm(S-2\alpha)$ & $\mp2$ & & & & \\
\cline{2-7}
 & $\pm(S-3\alpha)$ & $\mp3$ & \multirow{2}{*}{$\pm(S-2\alpha)$} & \multirow{2}{*}{$\mp2$} & \multirow{2}{*}{$\pm(S-2\alpha*)$} & \multirow{2}{*}{$\pm2$}\\
 \cline{2-3}
 & $\pm(S-4\alpha)$ & $\mp4$ & & & &\\
 \hline
\end{tabular}
\end{table}
We can summarize that for the emergent plateaus $\pm(S-\alpha,S-2\alpha,...,S-n\alpha...)$, Chern numbers are $(\mp1,\mp2...)$. These Chern numbers are equivalent to the winding numbers of corresponding edge states. The Chern numbers and the positions of magnetization plateaus can be unified in a generalized Streda formula \cite{Streda} for quasi-periodic spin-S chains: $C_{p}= \frac{\partial m_p}{\partial\alpha}$.

%\subsection{Butterfly-like spectrum}
{\bf Butterfly-like excitation spectrum.}
The incommensurate magnetization plateaus and non-trivial topological properties are totally determined by the modulation parameter $\alpha$ of quantum spin chains. In contrast to the adiabatical change of $\delta$, variation of $\alpha$ will change the structure of the excitation spectrum. To show the dependence of the spin-flip excitation spectrum on $\alpha\in(0,1)$, we calculate the spectrum with the variation of $\alpha$ under PBC for spin-$\frac{1}{2}$ and spin-1 systems in Fig.5(a) and Fig.5(b), respectively. It is amazing that the spin-flip excitation spectrum of the quasi-periodic Heisenberg model exhibits the similar structure of the Hofstadter's butterfly spectrum \cite{hofs}. The spectrum is mirror symmetric about $\alpha=\frac{1}{2}$ due to the invariance of Hamiltonian Eq.(\ref{xxx}) under transformation $(\alpha,\delta)\rightarrow(\alpha*\equiv 1-\alpha,-\delta)$. As $\alpha$ increases from $0$ to $\frac{1}{2}$, narrow bands merge into wider bands. For the spin-$\frac{1}{2}$ system, eventually three large bands and two large gaps survive. Quite different from the spin-$\frac{1}{2}$ case, there always exists a Haldane gap in the middle of the excitation spectrum for the spin-$1$ system, and finally six large bands are formed with the evolution of $\alpha$.
\begin{figure}
\includegraphics[width=3.8in]{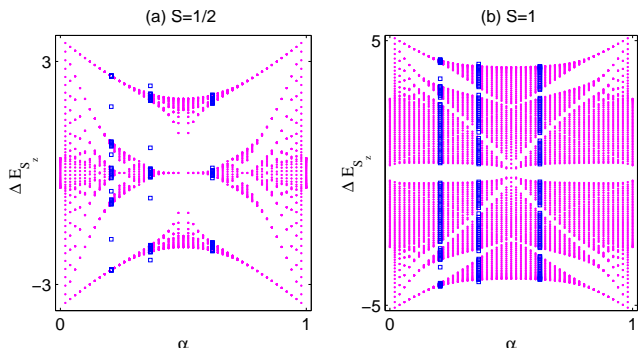}
\caption{(Color online) Spin-flip excitation spectrum versus $\alpha$ for (a) spin-$\frac{1}{2}$ and (b) spin-1 chains with $L=50$ and $\delta=0$ under PBC, where $\alpha$ is swept with step $0.02$ in interval $[0,1]$. The blue squares from left to right denote special cases with $\alpha=\frac{\sqrt{2}-1}{2}$, $\frac{\sqrt{3}-1}{2}$ and $\frac{\sqrt{5}-1}{2}$, respectively. Here we have taken $\lambda=0.8$ for (a) and $\lambda=0.3$ for (b).}
\end{figure}
Another notable feature is the existence of in-gap states once $L\alpha$ is not an integer under PBC. Excitation spectrum of quasi-periodic systems with irrational $\alpha$ always has in-gap states, which are the origin of slightly change of plateaus in the magnetization curves.
While no in-gap states exist for the periodic commensurate system with $\alpha L = integer$, the existence of in-gap states for the incommensurate chain under PBC is due to the mismatch of exchange coupling between the first and N-th site.
To see clearly the distribution of the in-gap state, we display real space profiles $\rho_{S_z}$ of plateau states at $m_l=-\frac{1}{2} + \frac{10}{50}$ and $m_u=-\frac{1}{2} + \frac{11}{50}$ for the system with $\alpha=\frac{\sqrt{2}-1}{2}$ and $L=50$ under PBC in Fig.6. As shown in the figure, both states exhibit similar distributions of incommensurate spin density waves with the main difference occurring around the mismatched bond between the final and first lattice site, which is clearly illustrated by the distribution of spin-flip excitations.
\begin{figure}
\includegraphics[width=3.8in]{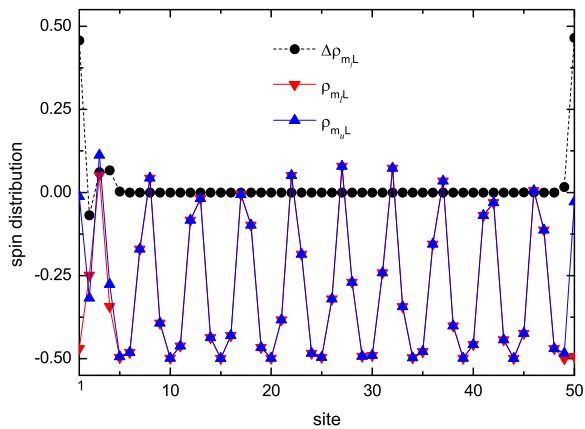}
\caption{(Color online) Real space profiles of spin density and spin-flip excitation for plateaus states of the spin-$\frac{1}{2}$ system with $\alpha=\frac{\sqrt{2}-1}{2}$, $\lambda=0.8$, $\delta=0$ and $L=50$ under PBC. Here $\rho_{m_l L}$ and $\rho_{m_u L}$ represent spin density distributions for states with $m_z=-\frac{3}{10}$ ($S_z = -15 $) and  $m_z=-\frac{7}{25}$ ($S_z = -14 $), respectively, whereas $\Delta \rho_{m_l L}=\rho_{m_u L} - \rho_{m_l L} $ represents the distribution of the spin-flip excitation of the plateau state $P_{\alpha}$.}
\end{figure}

\section*{Discussion}

By investigating the paradigmatic Heisenberg model with quasi-periodic geometry, we find a series of incommensurate magnetization plateaus. Under OBC, these non-trivial plateaus can host continuous edge states which connect the lower and upper excitation bands. The Chern numbers defined in a 2D parameter space to characterize different plateau states describe the winding patterns of edge modes. The topological properties of the magnetization plateaus are coded in a generalized Streda formula and the butterfly-like excitation spectrum. Our work unifies the quantum hall conductivity plateaus and quantized plateau states for quasi-periodic spin models in the scheme of strongly correlated topological insulators.

Our conclusion can be directly extended to the general XXZ spin models, with the spin exchange term $S_i^x S_{i+1}^x + S_i^y S_{i+1}^y + \Delta S_i^{z} S_{i+1}^z$. For AFM couplings, we find that the anisotropic exchange interactions do not destroy but stabilize these incommensurate plateaus as the plateau width has a positive correlation with $\Delta$ in the whole regime of $\Delta \geq 0$. The sweeping of the anisotropy parameter $\Delta$ produces a series of Hamiltonians which exhibit non-trivial $\alpha$-dependent magnetization plateaus. Particularly, for the quasi-periodic spin-$\frac{1}{2}$ chains, the XX model with $\Delta=0$ is exactly solvable and the non-trivial topological properties can be understood based on single-particle band theory of free fermions via a Jordan-Wigner transformation, where the latter can be exactly mapped to the famous 2D Hofstadter problem \cite{hofs}. The extension to the anisotropic high-spin system is straightforward and the general rules we summarized remain valid.

\section*{Methods}

The magnetization curves are determined by using the DMRG method which is the most powerful numerical tool for studying 1D strongly correlated systems. For the considered systems in the paper, the total $S_z$ is a good quantum number. Ground state energies in different subspace are compared to determine the magnetization under the specific magnetic field $h$.  Our DMRG simulations are rather reliable. The error truncation of the reduced density matrix is up to $10^{-8}$ to $10^{-12}$. We utilize four to fifteen sweeps to reach the convergence of the eighth digit for ground state energy per site. We have checked the accuracy of the DMRG algorithm by comparing the results from the exact diagonalization method on systems with lengths up to $L=24$.

The calculation of Chern numbers is settled in a 2D parameter space $(\theta,\delta)$. The Chern number is well defined for the ground state which is protected by a finite gap under PBC. Numerically, the continuous 2D space are divided into a discrete manifold \cite{tfu}. For $(\theta,\delta)\in[0,2\pi]\times[0,2\pi]$, we have analyzed different partitions: $5\times5$, $10\times10$, and $20\times20$ of the manifold and found the Chern numbers stay unchanged as long as we are considering the plateau states.

%\section*{References}

%\begin{addendum}

%\item [Acknowledgement]

\section*{Acknowledgment}
This work is supported by National Program for Basic Research of MOST(973 grant), NSFC under Grants No.11374354,
 No.11174360, No.11174115, No.10974234 and No.10934008, and PCSIRT (Grant No.IRT1251).

%\item [Author Contributions]

\section*{Author contributions}
H. P. H. and C. C. performed calculations. H. P. H., C. C. and S. C. analyzed numerical results. S. C. and H. G. L. planed and designed the study.
All contributed in completing the paper.
Correspondence and requests for materials should be addressed to Shu Chen (schen@iphy.ac.cn).

%\item [Competing Interests]
\section*{Author Information}
The authors declare that they have no competing financial interests.

%\end{addendum}

\end{document}